\documentclass[journal,draftcls,onecolumn,12pt,twoside]{IEEEtranTCOM}
\renewcommand{\baselinestretch}{1.6}
\usepackage{amsmath}
\usepackage{soul}
\usepackage{dcolumn}%
\usepackage{color}
\usepackage{filecontents,lipsum}
\usepackage[noadjust]{cite}
\usepackage{amsthm,amssymb,amsmath,bm}
\usepackage{subfigure}
\usepackage{amsfonts}
\usepackage{epsfig}
\usepackage{epstopdf}
\usepackage{amssymb}
\usepackage{amsmath}
\usepackage{cite}
\usepackage[Algorithm]{algorithm}
\usepackage{color,soul}
\usepackage{subfigure}
\usepackage{multirow}
\usepackage{rotating}
\usepackage{graphicx}
\usepackage{tabularx}
\usepackage{array}
\usepackage{color,soul}
\usepackage{graphicx,dblfloatfix}
\usepackage{blindtext}

\usepackage{amsthm,amssymb,amsmath,bm}
\usepackage{graphicx}
\usepackage{fancyhdr}
\usepackage{subfigure}
\usepackage[subfigure]{tocloft}
\usepackage[font={small}]{caption}
\usepackage{subfigure}
\usepackage{graphicx}
\usepackage{placeins}
\usepackage{float}
\usepackage{subfigure}
\usepackage{tabularx}
\usepackage{cite}
\usepackage{multicol}
\normalsize
\usepackage{booktabs}
\usepackage{calc, graphicx}  
\usepackage{epstopdf}  

\ifCLASSINFOpdf

\else

\fi
\normalsize

\begin{document}
	\title{End-to-End Performance Optimization in Hybrid Molecular and Electromagnetic Communications}

	%
	%
	%
	

	\author{Ali Momeni, Hamid Khoshfekr Rudsari,
		Mohammad Reza Javan, ~\IEEEmembership{Member,~IEEE}, Nader Mokari,~\IEEEmembership{Member,~IEEE}, Eduard A. Jorswieck,~\IEEEmembership{Senior Member,~IEEE}  and Mahdi Orooji,~\IEEEmembership{Member,~IEEE}
		\thanks{Ali Momeni is Research Assistant in Department	of Electrical and Computer Engineering, Tarbiat Modares University, Tehran
			14115111, Iran. }
		\thanks{Mohammad Reza Javan is with  Department of Electrical and Robotic Engineering, Shahrood University of Technology, Shahrud 3619995161, Iran.  }
		\thanks{Nader Mokari is with Department
			of Electrical and Computer Engineering, Tarbiat Modares University, Tehran
			14115111, Iran.  }
		\thanks{Eduard A. Jorswieck is with Department of Electrical Engineering and
			Information Technology, TU Dresden, Germany. }
		\thanks{Mahdi Orooji and Hamid Khoshfekr Rudsari are with Department of Biomedical Engineering, Tarbiat Modares University, Tehran
			14115111, Iran. }
	}
		

	%
	
\markboth{IEEE Transactions on Communications}%
{Submitted paper}

	\maketitle
	\vspace{-6em}
	\begin{abstract}
		
		Telemedicine refers to the use of information and communication technology to assist with medical information and services. In health care applications, high reliable communication links between the health care provider and the desired destination in the human body play a central role in designing end-to-end (E2E) telemedicine system. In the advanced health care applications,  $\text{e.g.}$ drug delivery, molecular communication becomes a major building block in bio-nano-medical applications. In this paper, an E2E communication link consisting of the electromagnetic and the molecular link is investigated. This paradigm is crucial when the body is a part of the communication system. Based on the quality of service (QoS) metrics, we present a closed-form expression for the E2E BER of the combination of molecular and wireless electromagnetic communications. \textcolor{black}{ Next, we formulate an optimization problem with the aim of minimizing the E2E BER of the system to achieve the optimal symbol duration for EC and DMC regarding the imposing delivery time from telemedicine services.} The proposed problem is solved by an iterative algorithm based on the bisection method.   Also, we study the impact of the system parameters, including drift velocity, detection threshold at the receiver in molecular communication, on the performance of the system. Numerical results show that the proposed method obtains the minimum E2E bit error probability by selecting an appropriate symbol duration of electromagnetic and molecular communications.

	\end{abstract}

	\begin{IEEEkeywords}
		Molecular communication, on-body communication, off-body communication, End-to-End Telemedicine.
		\vspace{-1em}
	\end{IEEEkeywords}
	\IEEEpeerreviewmaketitle
	\section{Introduction}
	
	In our modern era, the amount of the required health care services is rapidly increasing. These high expectancies are going to overtake a proportional increase in health system infrastructures and professional personnel \cite{weinstein2014telemedicine}. Telemedicine, which can be defined as the implementation of telecommunication technologies to provide medical services, has been introduced as a promising solution for the increasing shortages of the traditional medicine.
	
	 Telemedicine allows the measurement of biological and vital signals regardless of the borders and distances. In the
	 	telemedicine paradigm, the biological signals are obtained through variety of the biological sensing technologies, from the simple pulse oxiometer and temperature sensors to more sophisticated electrophysiological and electrocardiology  devices \cite{coiera2006communication}. The gathered signals must be post-processed, and eventually transmitted to the associated health care provider. The exterior communication hardwares are placed in order to form the pathway from the on-body devices to at a distant hospital or physicians’ station. One of the advanced applications of the telemedicine is the drug delivery which can be controlled via end-to-end (E2E) communication links. The drug delivery is an engineered method to deliver drugs to their targeted locations while minimizing the undesired side effects. One of the most important drug delivery applications is gene therapy. 
	 	The utilization of the drug delivery in the gene therapy allows to convey of the desired genetic information to the patient's organism \cite{blenke2016crispr}. The main challenge in the gene therapy is minimizing the risk of \emph{in vivo} toxicity and prolonging the lifespan of the payload. Also, the gene expression is short-lived due to the degradation of the plasmid in the nucleus \cite{chahibi2017molecular}. Consequently, the high reliable with minimum error probability E2E-telemedicine communication links are crucial in gene therapy.
	
	The E2E communication scenario plays a central and fundamental role in designing the telemedicine system between the health care provider and the human body. The applicability of the telemedicine crucially depends on the reliability of E2E communication links \cite{matusitz2007telemedicine}. Therefore, several quality of service (QoS) metrics are defined as the measurement of the performance criteria for E2E communication, in which several reliable communication links must work together in various circumstances and media including ``inner'', ``in-to-on'', ``on'', and ``off'' body areas.	
	
	Inspired by nature, one of the best solutions for inner body communication is to use chemical signals for carrying the information inside the human body to nanomachines through nanonetworks ~\cite{atakan2008channel,akyildiz2012nanonetworks}. This communication paradigm, which is called Molecular Communication (MC) \cite{farsad2016comprehensive}, has several advantages in comparison to electromagnetic based and acoustic wave based communication. The advantages includes but not limited to low energy consumption, the bio-compatible characteristics, and the existence of the biological receptors to serve as antenna \cite{farsad2016comprehensive}. Similar to Electromagnetic Communication (EC), different aspects of the communication have been studied in MC, such as channel modeling \cite{pierobon2010physical}, noise and interference \cite{pierobon2011diffusion}, and modulation and coding \cite{leeson2012forward}. Recently, Diffusion-based Molecular Communication (DMC) has been received a significant attention among various MC propagation scenarios such as walk-way or flow-based \cite{pierobon2010physical}. In DMC, without any additional and external energy, the molecules carrying the information propagate via Brownian motion in all available directions in a fluidic medium \cite{tiwari2016maximum,pierobon2010physical}.
 However, the major challenge of DMC is its high limited range of the communication \cite{llatser2013detection}. The intermediate nanomachines, which are serving as relays, are deployed to overcome this issue. The performance of relay-assisted DMC is investigated in~\cite{tavakkoli2017performance}.\
	
	In-to-on body (in2on-body) wireless communication is the EC between nanomachines inside the body and the wearable device on the surface of the body skin \cite{reusens2009characterization}. The main propagation environments of electromagnetic waves are inside and around the human body. On-body wireless communication is the interconnection and networking between wearable devices and the gateway transceivers \cite{reusens2009characterization}. 
	And ultimately, off-body wireless communication carries information from gateway transceivers to health care providers. The evaluation of the performance of the special scenario  involving both the ``in-body to on-body'' and ``on-body to off-body'' electromagnetic wireless propagation links, including bit error rate (BER), energy consumption, and data rate are considered in \cite{ntouni2014reliable}.
	
	
In this study, we assume that inner nanomachines send information to a distant health care provider via DMC, inner-body, in2on-body, on-body, and off-body links as E2E-telemedicine communication. The time is divided into multiple slots with equal durations. At each time slot, a message is transmitted from inner nanomachine to the distant health care provider.  In addition, the time slots are divided into several symbol durations for conveying the information in each link. The inner nanomachine sends its massage to the relay then the relay sends the received massage to the receiver nanomachine and ultimately through this sequence the massage is received by the health care system. Due to the fact that no buffer is considered in the nanomachines, EC and DMC communications should be in serial\footnote{ \textcolor{black}{It means as soon as  the packet arrives at an intermediate node, it is relayed over the next link towards the destination.}}. Therefore, BER and the delay of EC and DMC communication affects the E2E-BER and the total delay. It implies that the combination of EC and DMC must be considered as an integrated communication link. \textcolor{black}{In addition, some telemedicine services are imposing the limited delivery time of the command from the health care provider to the end nanomachine. \textcolor{black}{It results in a compromise between the performance of EC and DMC.}} The main question we aim to answer is: \emph{\textcolor{black}{How to choose the physical layer parameters, such as the symbol durations, in order to minimize the E2E BER?}} For answering this question, we firstly derive the analytical closed-form expression of E2E-BER when binary pulse shift keying (BPSK) modulation for EC and on-off keying (OOK) modulation for DMC is employed. Then, we formulate the optimization problem to determine the optimal symbol durations in EC and DMC and solve it by using the bisection algorithm. \textcolor{black}{It is important to emphasis that the main contribution of this paper is that what happen if we combine the EC and DMC as an integrated E2E-telemedicine communication and illustrate the trade off between the physical layer parameters such as symbol durations in EC and DMC which based on the best of authors' knowledge has not been addressed, yet.  } We also study the effect of the system parameters including the drift velocity and the detection threshold of DMC receiver on the performance of the E2E system. 
	
	The rest of paper is organized as follows: In Section \ref{Sec:System_Model}, the system model is described. In Section \ref{Sec:Channel Model}, the channel model of each communication is formulated, and the optimization problem is formulated and solved in Section \ref{Sec:Optimization}. The numerical results are presented in Section \ref{Sec:Results}, and finally, the paper is concluded in Section \ref{Sec:Conclusion}.

	\section{system model}	\label{Sec:System_Model}
\begin{figure}[t] 
		\centering
		\includegraphics[width=.61\textwidth]{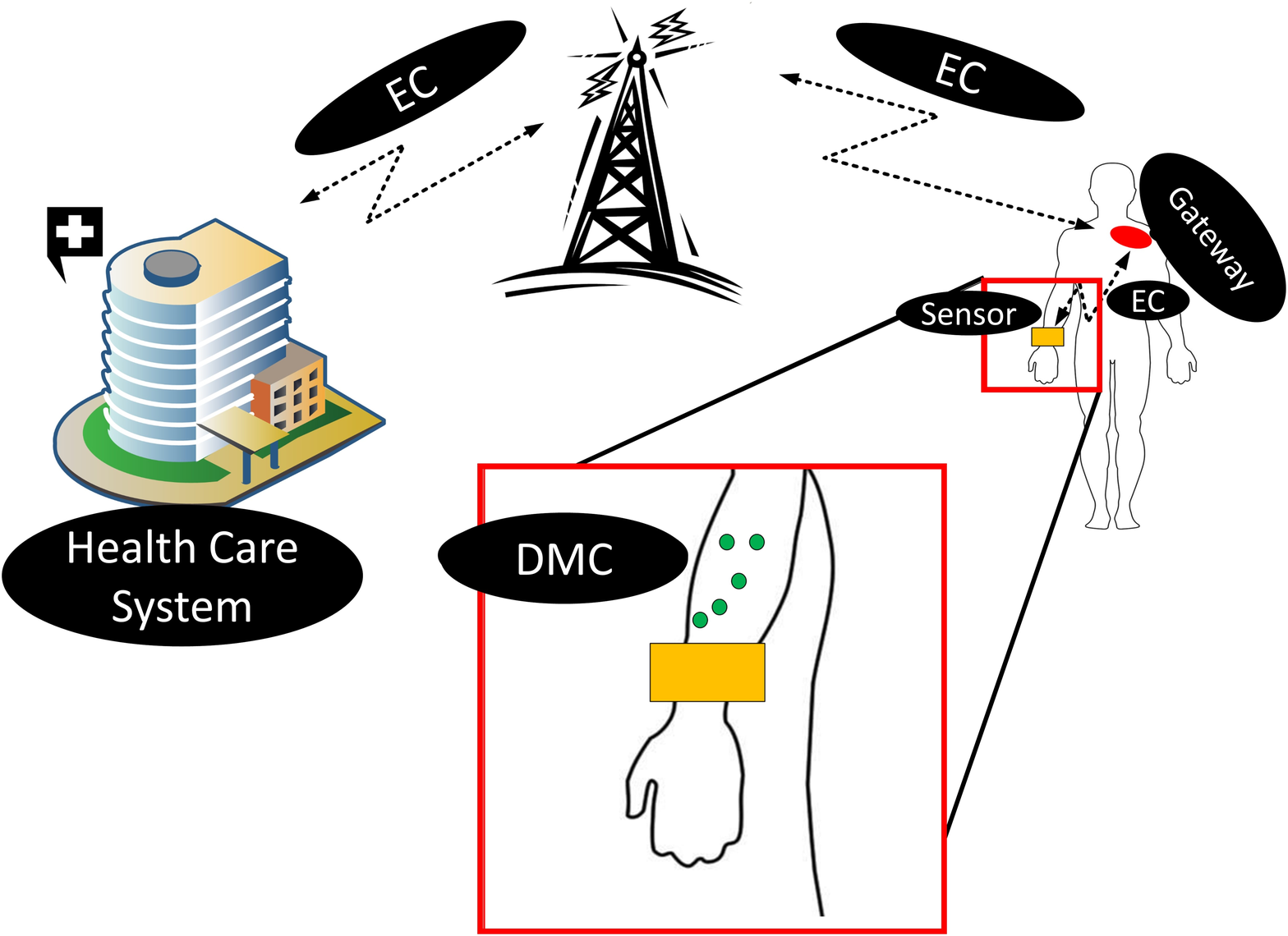}
		\caption{Overview of E2E e-health communication.}
		\label{Fig:Overal}
	\end{figure}

\begin{figure}[tb]
		\centering
		\includegraphics[width=.7\textwidth]{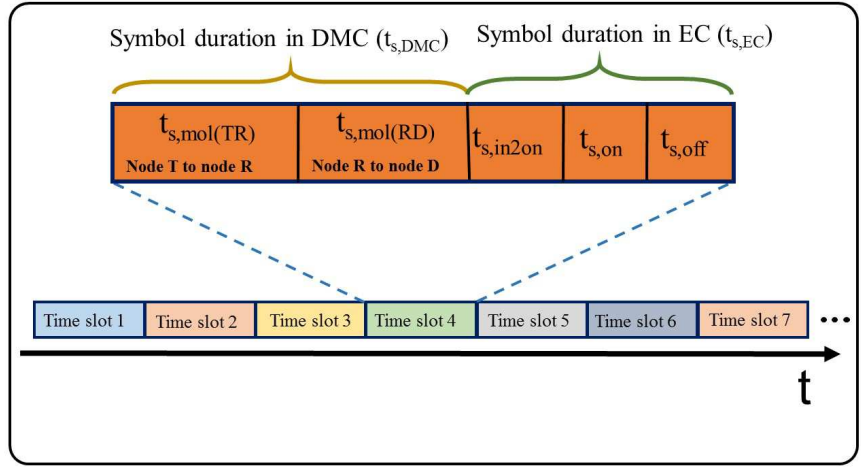}
		\caption{Symbol durations in the considered DMC and EC systems.}
		\label{Fig:Ts}
	\end{figure}

	We assume the E2E e-health communication includes molecular and electromagnetic wireless communication. In this E2E e-health system, the information is exchanged between the health care provider and nanomachines/receptors in the body environment. The considered E2E e-health system is shown in Fig.\ref{Fig:Overal}. One should note that all inner-body communications are DMC and in2on-body, on-body, and off-body communications are EC. \textcolor{black}{According to the imposed E2E delivery time, the total time duration of each time slot denoted by $t_\text{s}$, is divided into two time durations: one for DMC denoted by $t_\text{s,DMC}$, and the other one for EC denoted by $t_\text{s,EC}$  ($t_{\text{s}}=t_{\text{s,DMC}}+t_{\text{s,EC}}$).} Fig. \ref{Fig:Ts} illustrates the symbol time duration and the dedicated time intervals of each communication type in the proposed system model.   
	
	\subsection{Inner-body communication}
\begin{figure*}[t]
		\centering
		\includegraphics[width=.9\textwidth]{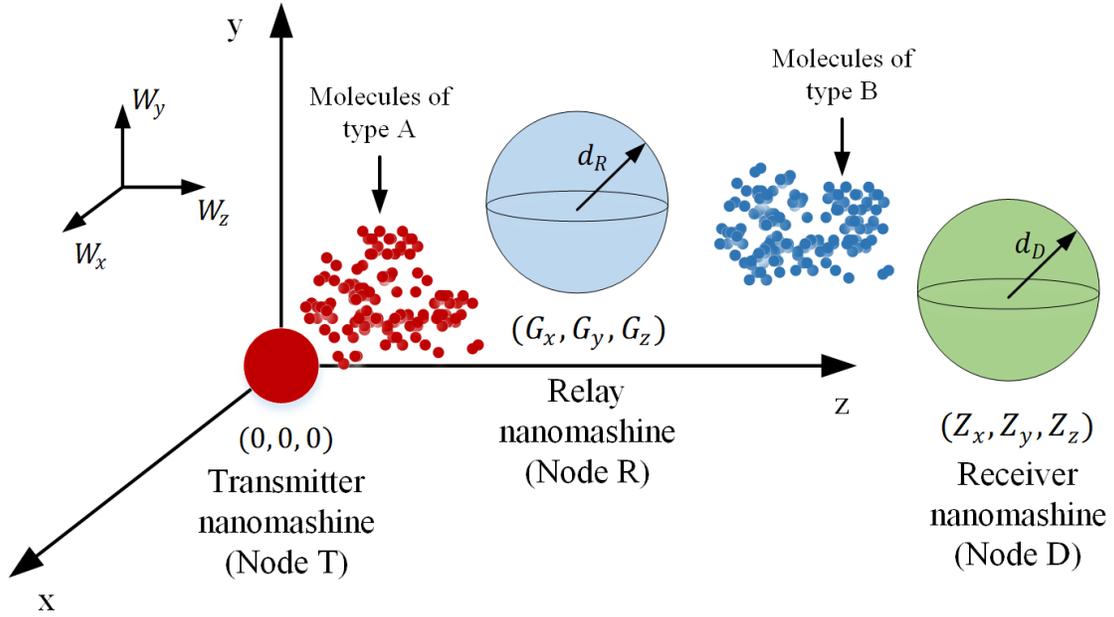}
		\caption{Relay-assisted diffusion-based molecular communication system.}
		\label{Fig:Relay}
	\end{figure*}

	 We use a generic nanotransmitter as the transmitter node, which can be reused as often as necessary. These types of transmitters are not natural and produced artificially \cite{chude2019advanced,fakhrullin2012cyborg}. The transmitter which resides in a liquid communication medium (for example blood), constitutes the transmitted signal by encoding the information onto the special type of the messenger molecules, and releases them to environment. Due to the very small sizes of messenger molecules (i.e., 1-10 nm in diameter), their propagation in this medium is governed by Brownian motion \cite{tepekule2015novel}. We use a relay-assisted DMC system, because such relay node can potentially improve the reliability and performance of a communication link \cite{ahmadzadeh2015analysis,einolghozati2013relaying,wang2015relay}.  The relay node resides in the environment can be an artificial nanomachine \cite{tavakkoli2017performance} or biological one \cite{ahmadzadeh2015analysis}. In nature, many biological systems have the both of molecule emission and reception capabilities \cite{atakan2008molecular,he2019chain}, for example, biochemical positive feedback loops process is common in protein channel of human body \cite{nakano2013molecular}. These are a key for designing a nanomachine, which acts as relay node.
     This MC system consists of a point source nanomachine, denoted by Node $T$, which only transmits information signals, a destination nanomachine, denoted by Node $D$, which receives molecular-type signals as information particles, and finally a relay nanomachine, denoted by Node $R$, which is relaying molecular-type signals. Fig. \ref{Fig:Relay} demonstrates the relay-assisted diffusion-based molecular communication system employed in this article.
	
	  As could be seen in Fig. \ref{Fig:Ts}, the relay-assisted molecular communication occurs at the beginning of each time slot, i.e., $t_\text{s,DMC}$. The assisting relay node is transmitting and receiving in the full-duplex fashion \cite{ahmadzadeh2015analysis}, and also we assume  $t_\text{s,DMC}$ is divided into two intervals of equal duration. In the first interval, denoted by $t_\text{s,mol(TR)}$, the transmitter node transmits the information to the relay. Next, the relay receives this information and retransmits the received information towards the destination in the second interval, denoted by $t_\text{s,mol(RD)}$. We adopt OOK modulation due to the fact that it is the most efficient binary modulation scheme in terms of molecular reception in DMC \cite{garralda2011diffusion}.
	  By using OOK modulation, the transmitter nodes ($T$ or $R$) release $Q$ molecules to send information bit ``1'' and no molecule to send information bit ``0'' at the beginning of the time slot\footnote{ In the case that another modulation scheme is employed in the molecular communication, just $\text{p}_{\text{mol}}^e$ in \eqref{Eq:Pe} and (\ref{eq:overall_prob}) should be adjusted accordingly.}. Also, node $T$ exploits type-$A$ molecules which can be detected by node $R$,  and node $R$ uses type-$B$ molecules which can be detected by node $D$. The use of two different types of molecules guarantees nearly interference free communication.

	\subsection{Wireless communications links}
	
	EC between nanomachines inside the body (for example in the blood vessels) and the wearable device on the body skin or at most 20 mm away from it, is called in2on-body communication \cite{ntouni2014reliable}. The wearable device communicates  with node $T$ electromagnetically in its own time interval. The communication between the wearable device on the body skin and the gateway on the body or at most 20 cm away from it, is called on-body communication \cite{ntouni2014reliable}. The gateway connects the on-body part to the off-body part via EC, in its own time interval. Finally, the off-body wireless part carries information from the gateway to the health care provider in the last time interval of the time slot. Further details of EC channel model and the corresponding BER are described in the following section. 
	
	The main parameters and notification used throughout this paper is listed in Table \ref{Table:Parameters}.

	\begin{table}[h]
		\caption{Description of terms and symbols used throughout this paper. }
		\label{Table:Parameters}
		\centering
		\begin{tabular}{@{}ll@{}}
			\toprule
			\textbf{parameter} &   \textbf{Variable} 
			\\ \midrule
			Drift velocity &   ($W_x$, $W_y$, $W_z$)  \\
			Diffusion coefficient &   D \\		Molecular noise mean &   $\mu_n$  \\
			Molecular noise variance &   $\sigma^2_n$   \\
			Molecular symbol duration &   $t_{\text{s,DMC}}$ \\
			Number of molecules for sending bit-1 &   $Q_A,Q_B$  \\
			Location of the relay node &   ($G_x$, $G_y$, $G_z$)   \\
			Location of the destination node & ($Z_x$, $Z_y$, $Z_z$)\\
			Detection threshold &   $\tau_{D}$     \\
			Distance between nanomachine and wearable device &   $d_{\text{in2on}}$   \\
			Distance between  wearable device and gateway &   $d_{\text{on}}$    \\
			SNR of in2on-body communication &   $\gamma _{\text{in2on}}$ \\
			SNR of on-body communication &   $\gamma _{\text{on}}$   \\
			SNR of off-body communication &   $\bar\gamma _{\text{off}}$   \\
			Total symbol duration time &   $t _{s,\text{total}}$  \\
			\bottomrule
		\end{tabular}
	\end{table}

	\section{E2E bit error performance }	\label{Sec:Channel Model}
	To derive the closed-form E2E-BER, each type of communication described in the previous section is modeled and ultimately, E2E-BER is calculated by concatenating them.  
	
	\subsection{Molecular communication channel model and BER}
	  We assume a diffusive environment, e.g., blood, inside the human body. There exists three nodes namely the transmitter node, relay node, and the receiver node in the introduced scheme. The transceivers are placed in the 3-dimensional (3-D) diffusive environment where the drift velocity is also attended. Therefore, the diffusion of the molecules are coalesced with the drift velocity of the environment to propagate them toward the receiver. The receiver considered in this paper is an spherical passive receiver which counts the number of molecules arrived in the volume of it, without absorbing them \cite{noel2016active}. The probability density function (pdf) of the molecule of type $\theta \in \{A,B\}$ released from the point source located at~(0,~0,~0) within time $t$ at which arrives at ($x$, $y$, $z$) is \cite{bhatnagar20193}
	  \begin{align}
	  	f(x,y,z,t) = \dfrac{1}{\sqrt{(4 \pi D_\theta t)^3}} \exp \bigg( -\dfrac{(x + G_x - W_x t)^2 + (y + G_y - W_y t)^2 + (z + G_z - W_z t)^2 }{4 D_\theta  t}  \bigg), \label{eq:pdf}
	  \end{align}
  	  where $D_\theta$, $\textbf{W} = (W_x, W_y, W_z)$, $\textbf{G} = (G_x, G_y, G_z)$, and $\exp$($\cdot$)  are the diffusion coefficient of the molecule of type $\theta$, the vector of the drift velocity of the medium, the location of the receiver, and the exponential function,  respectively. To attain the cumulative distribution function (CDF) of (\ref{eq:pdf}), i.e., the probability of arriving the molecules into the volume of the receiver, we should integrate (\ref{eq:pdf}) over the volume of the receiver. As stated in \cite{bhatnagar20193}, it could be approximated by using Simpson's rule. Therefore, the probability of arriving the molecules inside the volume of the spherical passive receiver with radius $d_R$ is
  	  \begin{align}
  	  F(t)  = \dfrac{d_R^2 }{144 \pi D_\theta t} \exp \bigg(\dfrac{- (G_z - W_z t)^2}{4 D_\theta t} \bigg) \cdot \bigg( 4 \sum_{k = 0}^{3} \lambda(k,t) +2 \sum_{k=1}^{3} \xi(k,t) + 2 \omega(t_s)   \bigg), \label{eq:cdf}
  	  \end{align}
  	  where $\lambda$, $\xi$, and $\omega$ are defined as follows \cite{bhatnagar20193}:
  	  \begin{align} 
  	  	\begin{split} \label{eq:sub_equation_CDF_alpha}
  	  	\lambda(k,t) =& \bigg[ \exp\bigg( -\dfrac{ (\frac{2k+1}{8} d_R + G_y - W_y t)^2}{4D_\theta t} \bigg)    +     \exp\bigg( -\dfrac{ (-\frac{2k+1}{8} d_R + G_y - W_y t)^2}{4D_\theta t} \bigg)\bigg]  \\& \times \bigg[ \text{erf} \bigg( \dfrac{\sqrt{1 - \dfrac{(2k + 1)^2}{64}} d_R - W_x t + G_x}{2 \sqrt{D_\theta t}}  \bigg)  -  \text{erf} \bigg( \dfrac{-\sqrt{1 - \dfrac{(2k + 1)^2}{64}} d_R - W_x t_s + G_x}{2 \sqrt{D_T t}}  \bigg)  \bigg],
  	  	\end{split} 
  	  	\\
  	  	\begin{split} \label{eq:sub_equation_CDF_beta}
  	  	\xi(k,t) =& \bigg[ \exp\bigg( -\dfrac{ (\frac{k}{4} d_R + G_y - W_y t_s)^2}{4D_\theta t} \bigg)    +     \exp\bigg( -\dfrac{ (-\frac{k}{4} r_R + G_y - W_y t)^2}{4D_\theta t} \bigg)\bigg]  \\& \times \bigg[ \text{erf} \bigg( \dfrac{\sqrt{1 - \dfrac{k^2}{16}} d_R - W_x t + G_x}{2 \sqrt{D_\theta t}}  \bigg)  -  \text{erf} \bigg( \dfrac{-\sqrt{1 - \dfrac{k^2}{16}} d_R - W_x t + G_x}{2 \sqrt{D_\theta t}}  \bigg)  \bigg],
  	  	\end{split}
  	  	\\
  	  	\begin{split} \label{eq:sub_equation_CDF_phi}
  	  	\omega(t) =& \exp \bigg(- \dfrac{(G_y - W_y t)^2}{4 D_\theta t} \bigg) \cdot \bigg[ \text{erf} \big( \dfrac{d_R - W_x t + G_x}{2 \sqrt{D_\theta t}} \big) - \text{erf} \big( \dfrac{-d_R - W_x t + G_x}{2 \sqrt{D_\theta t}} \big) \bigg].
  	  	\end{split}  		
  	  	\end{align}
    	We can write the CDF given in (\ref{eq:cdf}) as $F(t) = P_{\text{hit}}(\textbf{W},D_\theta,\textbf{G},t)$. As studied in \cite{rudsari2019non}, in case of considering $J$ relay nodes located at $(G_{1j}, G_{2j}, G_{3j})$ for $j~\in~\{ 1, 2, ..., J \} $, $(G_{1(j-1)} - G_{1(j)}, G_{2(j-1)} - G_{2(j)}, G_{3(j-1)} - G_{3(j)})$ is substituted by $(G_x, G_y, G_z)$ for calculating the CDF presented in (\ref{eq:cdf}).

	Node $R$ detects the type $A$ molecules that are transmitted by node $T$. In the receiver of node $R$, maximum-a-posterior probability (MAP) rule is employed for detection of the transmitted molecule \cite{tavakkoli2017performance}. The information bit detected by node $R$, denoted by $x_R \in \{0,1\}$, is given by 

	\begin{align}
		{x_R} = \left\{ {\begin{array}{*{20}{c}}
				{1,{\rm{   \: \:\:\:  \:   \:\:\:  if\: g}}_{T,R}^A \ge {\tau _R},}\\
				{0,{\rm{   \: \:\:\:\:  \:\:   \: if\: g}}_{T,R}^A < {\tau _R},}
		\end{array}} \right.
		\label{Eq:MAP}
	\end{align}
	where $g_{T,R}^A $ is the total number of $A$ molecules absorbed by node $R$ in the source transmission time interval, i.e., $t_{\text{s,mol(TR)}}$, and $ \tau_R $ is the detection threshold at node $R$. After decoding, node $R$ re-encodes $x_R$ and forwards it to node $D$ in the relay transmission time interval, i.e., $t_{\text{\text{s,mol(RD)}}}$. The information bit detected by node $D$ is denoted by
	$x_D$. $ g_{T,R}^A$ follows the normal distribution as \cite{tavakkoli2017performance}
	\begin{align}
		{\Pr}\bigg(g_{T,R}^\theta ~\bigg|~ {x_T} = 0\bigg) &\sim \mathcal{N} \bigg({{\mu }_{{{0}_{T,R}}}},{\sigma}_{{0}_{T,R}}^2\bigg),\\
		{\Pr}\bigg(g_{T,R}^\theta ~\bigg|~ {x_T} = 1\bigg) &\sim \mathcal{N}\bigg({\mu _{{1}_{T,R}}},\sigma _{{1}_{T,R}}^2\bigg),
	\end{align}
	where $\mathcal{N}(.,.)$ is the normal distribution, ${{\mu }_{{{0}_{T,R}}}}$ and ${{\mu }_{{{1}_{T,R}}}}$ are mean values and ${\sigma}_{{0}_{T,R}}^2$ and ${\sigma}_{{1}_{T,R}}^2$ are variance values which are derived as
	follows \footnote{In this paper, the intersymbol interference is ignored. The analysis of this interference is out of the scope and is relegated as future works.} \cite{tavakkoli2017performance}:
\begin{align}
{{{\mu }_{{{0}_{T,R}}}}} =& 0.5{Q_A} {q_{T,R}^A}  + {\mu _{n}},\\
{{{\mu }_{{{1}_{T,R}}}}} =& 0.5{Q_A}{q_{T,R}^A}  + {Q_A}{P_{1T,R}^A}+{\mu _{n}},\\
{{\sigma }_{{{0}_{T,R}}}}^2 =& 0.5{Q_A}{q_{T,R}^A} (1 - q_{T,R}^A) + 0.25Q_A^2{(q_{T,R}^A} {)^2}+ \sigma _{n}^2 + {\mu _{n}},\\
{{\sigma }_{{{1}_{T,R}}}}^2 =& {Q_A}P_{1T,R}^A(1 - P_{1T,R}^A) + 0.5{Q_A} {q_{T,R}^A} (1 - q_{T,R}^A)+ 0.25Q_A^2  {(q_{T,R}^A} {)^2} + \sigma _{n}^2 + {\mu _{1T,R}},
	\end{align}
where ${q_{T,R}^A}=P_{1T,R}^A$ in which $P_{1T,R}^A=P_{\text{hit}}(\textbf{W},D_A,\textbf{G},t_{\text{s,DMC}}/2)$, and $t_{\text{s,DMC}}$ is symbol duration in DMC, and $d_{t,d}$ is the distance between nodes T and D. The value of $\Pr(x_R= 1)$ can be calculated as \cite{tavakkoli2017performance}	
			\begin{align} 		
{\Pr} \bigg({x_R} = 1\bigg)=\Pr \bigg(x_T= 1 \bigg){\Pr}\bigg({{ x}_R} = 1 ~\bigg|~ {x_T}= 1\bigg)+ \Pr \bigg(x_T= 0 \bigg){\Pr}\bigg({{ x}_R} = 1 ~\bigg|~ {x_T} = 0\bigg).
		\label{Eq:Prob_Xr1}
	\end{align}
	
	 By using the MAP detection rule in \eqref{Eq:MAP}, the conditional probabilities in (10) can be written as \cite{singhal2015performance}
	\begin{align}
		{\Pr}\bigg({{ x}_R} = 1 ~\bigg|~ {x_T} = {s}\bigg)=  {\Pr}\bigg(g_{T,R}^\theta  \ge {\tau _R} ~\bigg|~ {x_T} = {s}\bigg)
		= \frac{1}{2}\left(1 - \text{erf}\left(\frac{{{\tau _R} - {{\mu }_{{{s}_{T,R}}}}}}{{\sqrt {2{{\sigma }_{{{s}_{T,R}}}}^2} }}\right)\right),
	\end{align}
	where $\text{erf}(\cdot)$ is the error function and $s\in\{0,1\}$. In this model, the communication occurs with no error if $ x_D=x_T$ holds. Consequently, the error probability for the $n$th bit is calculated as
	\begin{align} 	\label{Eq:P_e_mol}
		P_{\text{mol}}^e = {\Pr \big({x_T}= {0}\big)} \Pr \bigg({x_D} = {1}~ \bigg	|~ {x_T}= {0}\bigg)
		+{\Pr \big({x_T}= {1}\big)} \Pr \bigg({x_D} = {0}~ \bigg	|~ {x_T}= {1}\bigg).
	\end{align}
	By exploiting the chain rule, the first term on the right side of (12) can be obtained as follows  
	\begin{align} \label{Eq:Pr_S0}
\Pr \bigg({ x_D} = 0 ~\bigg|~{x_T}= 1\bigg) &=\Pr \bigg({x_R} = 0~\bigg|~ {x_T} = 1\bigg) \times \Pr \bigg({ x_D} = 0 ~\bigg|~{x_R}= 0,{x_T}= 1\bigg)\\\nonumber &+ \Pr \bigg({x_R}= 1 ~\bigg|~ {x_T}= 1\bigg) \times \Pr \bigg({ x_D}= 0 ~\bigg|~{x_R} = 1,{x_T} = 1\bigg).\
	\end{align}

A similar approach can be employed to extend the second term of (12). Finally, after some manipulations, the error probability in \eqref{Eq:P_e_mol} can be written as
	\begin{align}
		P_{\text{mol}}^e = \frac{1}{2} &+ \frac{1}{8}\left[\text{erf}\left(\frac{{{\tau _R} - {\mu _{1T,R}}}}{{\sqrt {2\sigma _{1s,r}^2} }}\right) - \text{erf}\left(\frac{{{\tau _R} - {\mu _{0T,R}}}}{{\sqrt {2\sigma _{0T,R}^2} }}\right)\right] \nonumber\\
		&\times \left[\text{erf}\left(\frac{{{\tau _D} - {\mu _{0R,D}}}}{{\sqrt {2\sigma _{0R,D}^2} }}\right) - \text{erf}\left(\frac{{{\tau _D} - {\mu _{1R,D}}}}{{\sqrt {2\sigma _{1R,D}^2} }}\right)\right],\label{Eq:Pe}
	\end{align}
where $\tau_D$ denotes the detection threshold at node $D$.

\subsection{In2on body communication channel model and BER}
Our goal in this subsection is to model the statistical BER of the in2on-body channel. The SNR of the link between the nanomachine and the wearable device can be expressed as
	\begin{align}
		{\gamma _{\text{in2on}}} = \frac{{
				\vartheta t_{s,\text{in2on}} {P_\text{in2on}}}}{{{N_0}}}{10^{-\frac{b}{{10}}}},
	\end{align}

\begin{align}
\vartheta  = \exp\left(-\frac{{{P_L}({d_0}) + 10n{{\log }_{10}}\frac{d_\text{in2on}}{{{d_0}}}}}{{10}} \ln 10 \right),
\end{align}
where ${P_L}({d_0})$ is the path loss in dB at a reference distance $d_0$, $P_\text{in2on}$ is the transmission power of the nanomachine, $t_{s,\text{in2on}}$ is the symbol duration of in2on body channel, $n$ is the path loss exponent, $d_\text{in2on}$ is the distance between nanomachine and the wearable device, $N_0$ is the power spectral density of the zero mean complex additive white Gaussian noise, and $b\sim \mathcal N(0,\sigma^2)$ is a normally distributed random variable that models shadowing effect \cite{yazdandoost2007channel,ntouni2014reliable}.

\textcolor{black}{It is clear that, the above parameters corresponding to in2on body communication depend on location of nanomachines inside the blood vessel in deep tissue or near-surface of skin which details of the model derivation can be found in \cite{miniutti2008narrowband} and are summarized  in table IV}

	By considering Binary Phase-Shift Keying (BPSK) modulation, the BER of the instantaneous SNR (denoted by  $\gamma _{\text{in2on}}$) is given by
	\begin{align}
		P_{\text{in2on}}^e(\gamma _{\text{in2on}} )= \frac{1}{2}\text{erfc}(\sqrt{ \gamma _{\text{in2on}}}),
		\label{Eq:P_in2on}
	\end{align}
	where erfc($\cdot$) is the complementary error function. The Average BER (ABER) is derived by averaging  $P_{\text{in2on}}^e$ over $\gamma$ which is given by
	\begin{align}
	 \bar{P}_{{\rm{in2on}}}^{\rm{e}} = \int\limits_0^\infty  {P_{{\rm{in2on}}}^{\rm{e}}(\gamma )} {f_\gamma }(\gamma ){\rm{d}}\gamma.
	 \end{align}
	 Considering the fact that the SNR is log-normally distributed, the ABER in (18) does not have the analytical closed-form solution \cite{ntouni2014reliable}. Therefore, the complementary error function is approximated as follows \cite{loskot2009prony}:

	\begin{align} \label{Eq:log_norm_app}
		&\text{erfc}(x) = 2\bigg(0.168{e^{ - 1.752{x^2}}} + 0.144{e^{ - 1.05{x^2}}} + 0.002{e^{ - 1.206{x^2}}}\bigg).
	\end{align}
	
	Therefore, the ABER for the log-normal model, using the approximation in \eqref{Eq:log_norm_app}, is expressed as 
	
	\begin{align}
		\bar{P}_{\text{in2on}}^e = 0.168 \Phi\bigg(1.752{e^{{\mu _\gamma } + \frac{{\sigma _\gamma ^2}}{2}}},\frac{{{\sigma _\gamma }}}{2}\bigg)+
		 0.144\Phi\bigg(1.05{e^{{\mu _\gamma } + \frac{{\sigma _\gamma ^2}}{2}}},\frac{{{\sigma _\gamma }}}{2}\bigg) + 0.002\Phi\bigg(1.206{e^{{\mu _\gamma } + \frac{{\sigma _\gamma ^2}}{2}}},\frac{{{\sigma _\gamma }}}{2}\bigg),
	\end{align}
	where $\mu_\gamma=\text{ln}(\vartheta  t_{s,\text{in2on}}{P_{\text{in2on}}}/{N_0})$, $\sigma_\gamma={\sigma_{\text{in2on}}\text{ln}10}/{10}$ and $\Phi(.)$ is the Frustration function defined by ~\cite{abou2011cooperative}
	\begin{align}
		\Phi(k,l) = \int\limits_0^\infty  {\frac{1}{{x\sqrt {2\pi } l}}} {e^{ - k{x^2}}}{e^{ - \frac{{{{(\ln (x) + {l^2})}^2}}}{{2{l^2}}}}}dx.
	\end{align}

	\subsection{On-body communication channel model and BER}
	
	The path loss of the link between the wearable device and the gateway is a function of the distance, the part of the body that the wearable device is located, and the motion of the human body \cite{ntouni2014reliable}. The best fitting distribution for these scenarios has been found to be the log-normal distribution \cite{yazdandoost2007channel}. 
	The closed-form of BER for the case that SNR is log-normally distributed with mean $\mu$, and scale parameter $\sigma$, can be obtained in similar to (20).
	For the case that SNR is log-normally distributed with mean $\mu$, and scale parameter $\sigma$,  the BER can be solved in the closed-form similar to (20).
	However, in this scenario, we have $\mu_\gamma=\mu+\text{ln}(\gamma _{\text{on}})$, $\sigma_\gamma=\sigma_{\text{on}}$ and $\gamma _{\text{on}}=({P_{\text{loss}}P_wt_{\text{s,on}}})/{N_0}$, where $P_w$ and $t_{s,\text{on}}$ are the transmit power of the wearable device, and the symbol duration time of on-body channel, respectively.

	\subsection{Off-body communication channel model and BER}
	 Rayleigh fading channel with additive white Gaussian noise is assumed for off-body communication link. Thus, average BER of BPSK is given by \cite{goldsmith2005wireless}
	\begin{equation}
		\bar P_{_{{\rm{\text{off}}}}}^e = \frac{1}{2}\bigg[1 - \sqrt {\frac{{{{\bar \gamma }_{\text{off} }}}}{{1 + {{\bar \gamma }_{\text{off}}}}}}\bigg ],
	\end{equation}
	where  ${\bar \gamma _{\text{off}}} = ({{{P_g}}~t_{\text{s,off}}~d_{\text{off}}~^{-\alpha}})/{{{N_0}}}$  is SNR per bit. $P_g$, $t_{\text{s,off}}$, $d_{\text{off}}$, and $\alpha$ are transmit power of the gate-way device, symbol duration of off-body channel, the distance between the gate-way and the access point and the path loss exponent, respectively.

	\subsection{E2E communication channel model and BER}
	
	In order to compute the E2E-BER, we use the fact that BERs  of the above-mentioned communication links are independent, due to their independent physical mediums. Therefore, the total bit error probability of E2E system can be written as
	\textcolor{black}{
	\begin{align}
		P_{\text{E2E}}^e = \sum_{q \in Y} P_q^e \prod_{k \in Y, k \neq q} P_k^c 
	+ \sum_{q \in Y} P_q^c \prod_{k \in Y, k \neq q} P_k^e , \label{eq:overall_prob}
	\end{align}}where $P_{\text{E2E}}^e$ is the total bit error probability and $P_{q}^e$ and $P_{q}^c$ are the bit error probability and bit correct probability of link $Y\in \{ \text{mol},{\rm{ \text{in2on}, on, \text{off}\} }}$, respectively. In addition, $P_q^c = 1 - P_q^e\ $.
	
	\section{\textcolor{black}{symbol duration ratio optimization problem}} \label{Sec:Optimization} 

	As mentioned in section II, the total time duration of each time slot denoted by $t_\text{s}$, is divided into two time durations: one for DMC denoted by $t_\text{s,DMC}$, and the other one for EC denoted by $t_\text{s,EC}$. The time duration of each slot for E2E communication is predetermined and set to $t_\text{s}$ which implies the following equation holds:	
	\begin{equation}
	t_{s}=t_{\text{s,DMC}}+t_{\text{s,EC}},
	\end{equation} 
	where $t_{\text{s,EC}}$ indicates symbol duration of the entire wireless communications including on-body, in2on-body, and off-body communication.\
	The important question is that how $t_s$ is chosen? In fact, some telemedicine services are imposing the limited delivery time of the command from the health care provider to the end nanomachine. For example  the gene therapy drug delivery  allows us to transfer the desired genetic information to the patient's organism \cite{blenke2016crispr}. In this E2E-telemedicine case, it is important to convey the genetic information with minimum risk and at a specified time due to performing the desired chemical reactions \cite{chahibi2017molecular}. Consequently, the value of $t_s$ is imposed by telemedicine services regarding the type of drug delivery process. Consequently, the value of $t_s$ is imposed by telemedicine services regarding the type of drug delivery process. By fixing $t_s$, this predefined delay time is divided into EC and DMC which leads to a compromise between the performance of the EC and DMC. Once again, we emphasize that  $t_s$ is fix and predetermined via telemedicine services \cite{chahibi2017molecular,felicetti2016applications}. For example, in case of utilizing DMC in drug delivery, the time of releasing the drugs into the intended location is controlled by the drug releasing mechanism \cite{chahibi2017molecular}.
	 Our target is finding the symbol duration ratio ($t_{s,\text{DMC}}$) in a way that (24) holds.
	Next, we formulate an optimization algorithm for finding the optimum time slot partitioning to achieve the best performance in E2E communication.
	The objective is to minimize the E2E BER of the
	system. Therefore, the problem
	is formulated as	
	\begin{align}
		\begin{array}{l}
			\mathop {\min }\limits_{{t_{\text{s,DMC}}}} \, \, \,  {\rm{P}}_{\text{E2E}}^e\\
			\text{s.t} \, \, \, \, \, \, \, \, \, \,	t_{s}=t_{\text{s,DMC}}+t_{\text{s,EC}}.
		\end{array}\
				\label{Eq:min}
	\end{align}

	\begin{table}[t]
	\caption{PROPOSED ALGORITHM BASED ON THE BISECTION METHOD}
	\label{my-label}
	\centering
	\begin{tabular}{@{}l@{}}
		
		\\ \midrule
		\textbf{Initialization}: \\
		\, \, \,	Set lower-bound= 0, upper-bound= 1 and $0<\epsilon<1$\\
		\textbf{Iterations}:\\
		\textbf{Step 1}: h= (Lower-bound + upper-bound)/2.\\
		\textbf{Step 2}: Solve the convex feasibility problem (27).\\
		\textbf{Step 3}: If (27) is feasible, upper-bound= h; else lower-bound= h.\\
		\textbf{Step 4}: If $|$upper-bound - lower-bound$|$ $\le$ $\epsilon$, then stop. Otherwise, go back to Step 1.
		\\
		\bottomrule
	\end{tabular}
\end{table}
	 We also assume  $t_\text{s,EC}$ is divided into three intervals of equal duration associated to in2on-body, on-body and off-body communication links.
	One should note that the objective function in \eqref{Eq:min} is not a convex function, and hence, the
	optimization problem is not convex. However, it could be shown that the objective function is quasiconvex (see Fig.~\ref{Fig:BER_vs_MAX}), due to the fact that its domain and all its sublevel sets are convex \cite{boyd2004convex,tavakkoli2017performance}.
	One method to solve the
	quasiconvex optimization problem relies on the representation
	of the sublevel sets of a quasiconvex function via a family
	of convex inequalities, as described in \cite{boyd2004convex}. For solving the optimization problem \eqref{Eq:min},  we utilized the bisection method, in which, the optimal symbol duration is computed by solving a convex feasibility problem at each step.  
	According to \cite{boyd2004convex}, the $h$-sublevel set ($h \in\mathbb{R}$) of a function $f$ : $\mathbb{R}^n\to \mathbb{R}$ is defined as 
			\begin{equation}
		{\psi _h} = \{ x \in {\rm{\textbf{dom} }}~f~|~f(x) \le h\} ,
		\end{equation}
	where $\mathbb{R}$ is the set of real numbers.
	Consequently, the new optimization problem whose constraint is the $h$-sublevel set of the objective function in problem \eqref{Eq:min}
	can be written as	
	\begin{align}
		\text{Find}  \, \, \, \, \, \,&t_{\text{s,DMC}} \\ \nonumber 
		\text{s.t}~~~~ &{\rm{P}}_{\text{E2E}}^e-h<0,\\ \nonumber
		&t_{s}=t_{\text{s,DMC}}+t_{\text{s,EC}}.\nonumber\\\nonumber            
	\end{align}
The proposed algorithm based on the bisection method for solving our problem is presented in Table II where $\epsilon$
	is a positive small number.

	The bisection optimization solution is a function of the DMC parameters and SNR of EC where SNR of the EC is estimated at the beginning of the transmission as a transmission routine procedure, and DMC parameters are pre-determined and fixed. Therefore, the bisection algorithm is run in the EC transmission side which does not have computational complexities. On the other hand, from the computational point of view, this method is a simple and robust root-finding mathematical algorithm which is of low complexity as investigated in \cite{sikorski1982bisection}.

\begin{center}
		\begin{table}[tb]
		\caption{Simulation setting.}
		\label{my-label}
	\centering	
			\begin{tabular}{@{}llll@{}}
				\toprule
				\textbf{Variable} & \textbf{values} &\textbf{Variable} & \textbf{values} \\ \midrule
				$W_x, W_y, W_z$ \cite{felicetti2014modeling} & [1, 100] $\mu \text{m/s}$ &
				D \cite{srinivas2012molecular} &  $\text{4}\times \text{10}^{\text{-9}}$~m$^2$/s\\	
				$\mu_n$ \cite{srinivas2012molecular} & 40 &
				$\sigma^2_n$ \cite{srinivas2012molecular}  & 100  \\
				$t_{\text{s,DMC}}$ & {[}2-7{]} ms &
				$Q_A,Q_B$  & {[}100-1000{]} \\
				$Z_x, Z_y, Z_z$ \cite{bhatnagar20193}  & [20-200] $\mu$m  &
				$d_R, d_D$  & 100~$\mu\text{m}$   \\
				$d_{\text{in2on}}$  & 20 mm  &
				$d_{\text{on}}$  & 1.3 m  \\
				$t _{s,\text{total}}$ & 9 ms & 
				 $P_{\text{loss}}$ \cite{ntouni2014reliable} & 63.24 dBm \\  
				$\mu$ \cite{ntouni2014reliable} &  -0.39& 
				$\sigma_{\text{on}}$ \cite{ntouni2014reliable} & 0.23 \\ 
				\bottomrule
			\end{tabular}
	\end{table}
\end{center}

\begin{center}
	\begin{table}[tb]
		\caption{In2on body communication link parameters.}
		\label{my-label}
		\centering	
		\begin{tabular}{@{}lll@{}}
			\toprule
			\textbf{Variable} & \textbf{Deep tissue} &\textbf{Near surface }  \\ \midrule
			$P_L(d_0)$ &47.14  &
			49.81  \\
				$n$ &4.26  &
			4.22 \\	
				$\sigma_{\text{in2on}}$ &7.85  &
			6.81 \\			\bottomrule
		\end{tabular}
	\end{table}
\end{center}

	\section{Numerical Results} \label{Sec:Results}
	
	In this section, we present the numerical
	results to evaluate the error probability performance of the
	proposed E2E communication system. We also show how the system parameters affect the
	performance. We consider a diffusive environment like blood, where the molecules propagated through it. The presumed environment has drift velocity in three dimensions, which would be a precise  model for turbulence of such environment like blood, pheromones in air, and toxic chemicals \cite{bhatnagar20193}.   
	In molecular communication, we assume node R is placed between nodes T and D, and it emits the same number of molecules that node T transmits to relay the information\footnote{Without loss of generality, no channel coding is considered for EC and DMC. If a particular channel coding is engaged, the probability of bit errors must be adjusted accordingly}. It is also assumed that the released molecules of type A and type B have the same diffusion coefficient in the
	medium. The system parameters used in the analysis and simulations are given in Table III.
	\begin{figure}[]
		\centering
		\scalebox{.1}{}
		\includegraphics[width=330pt]{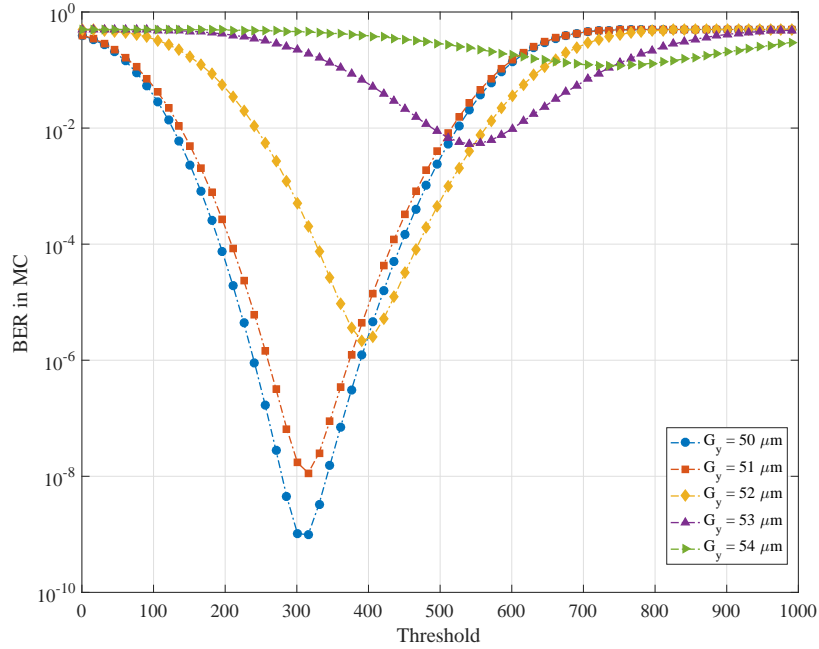}
		\caption{The molecular BER performance as a function of different values of threshold at the receiver for different values of the relay nodes' location ($W_x = \text{10}\mu\text{m/s}$, $W_y = \text{30} \mu\text{m/s}$, $W_z = \text{20}~\mu\text{m/s}$, $Z_x = \text{200}~\mu\text{m} = 2~G_x$, $Z_y =\text{100}~\mu\text{m}$, $Z_z = \text{20}~\mu\text{m} = 2~G_z$, $t_{s,\text{DMC}} = 4~\text{m}s$, and $Q_A = Q_B = 1000$) }
		\label{Fig:BER_vs_threhold}
	\end{figure}
	
	Choosing an appropriate value for threshold at the destination node can improve the performance of such DMC system \cite{ahmadzadeh2015analysis,einolghozati2013relaying}. Therefore, the performance  of the error probability in DMC is analyzed in Fig.~\ref{Fig:BER_vs_threhold}. In this figure, the BER of DMC as a function of threshold at the receiver for different values of relay location is shown. It shows that by changing the location of node R, the BER is also changed. The location of the destination node in y axis is $\text{100}~\mu\text{m}$. By setting the relay location in y axis as  $\text{50}~\mu\text{m}$ where is the middle location between nodes S and D, the BER achieves the best performance. However, the closer the relay node to the destination node, the less the BER performance in DMC. By increasing the location of node R in y axis from $\text{50}~\mu\text{m}$ with 300 molecules as the threshold, to $\text{54}~\mu\text{m}$ with 750 molecules as the threshold, the BER  decreases from around $\text{10}^{\text{-9}}$ to around $\text{10}^{\text{-1}}$.
	\begin{figure}[]
		\centering
		\scalebox{.1}{}
		\includegraphics[width=330pt]{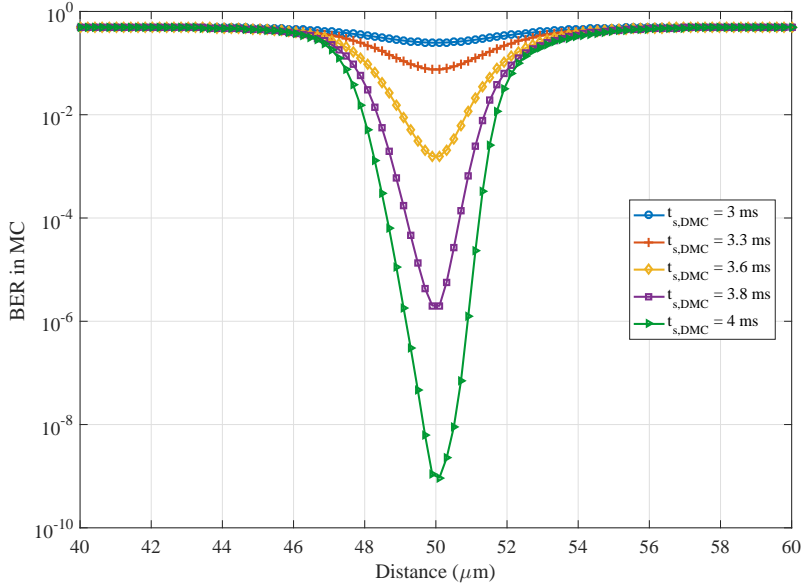}
		\caption{The molecular BER performance as a function of different values of $G_y$ for different values of symbol duration ($W_x = \text{10}\mu\text{m/s}$, $W_y = \text{30} \mu\text{m/s}$, $W_z = \text{20}~\mu\text{m/s}$, $Z_x = \text{200}~\mu\text{m} = 2~G_x$, $Z_y =\text{100}~\mu\text{m}$, $Z_z = \text{20}~\mu\text{m} = 2~G_z$, and $Q_A = Q_B = 1000$). }
		\label{Fig:BER_vs_distance}
	\end{figure}
	
	In addition, the BER in DMC as a function of $G_y$ (relay location in y axis) for different values of symbol duration in DMC ($t_{s,\text{DMC}}$) is shown in  Fig.~\ref{Fig:BER_vs_distance}. It could be comprehended that by increasing $t_{s,DMC}$ from 3~ms to 4~ms, the BER in DMC can be improved. It is also shows that BER can be improved by locating the relay node at the middle of the transmitter and the destination node.
		\begin{figure}[t]
		\centering
		\scalebox{.1}{}
		\includegraphics[width=330pt]{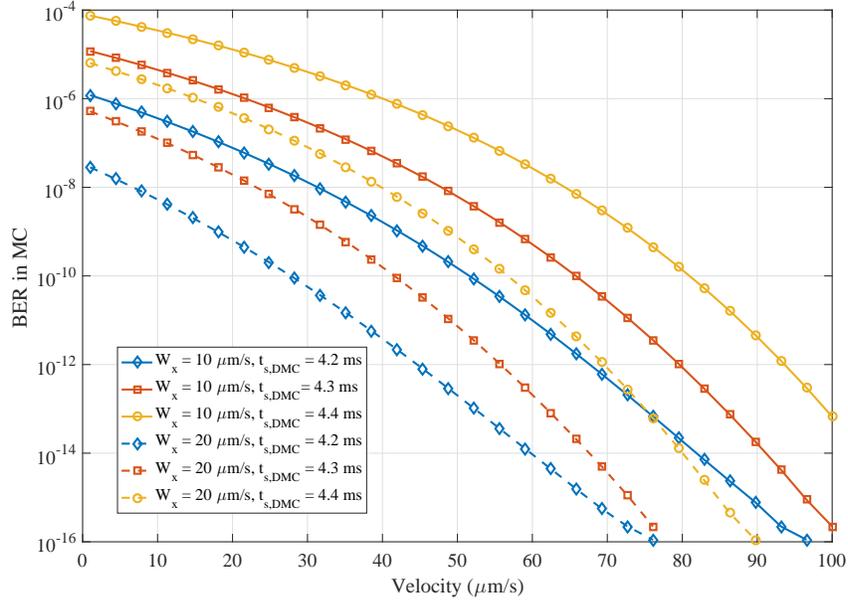}
		\caption{The molecular BER performance as a function of different values of $W_y$ for different values of $W_x$ and symbol duration ($W_z = \text{20}~\mu\text{m/s}$, $Z_x = \text{200}~\mu\text{m} = 2~G_x$, $Z_y =\text{100}~\mu\text{m} = 2~G_y$, $Z_z = \text{20}~\mu\text{m} = 2~G_z$, and $Q_A = Q_B = 1000$) }
		\label{Fig:BER_vs_velocity}
	\end{figure}
	
	The BER in DMC is also analyzed as a function of the drift velocity in y axis of the environment for different values of $W_x$ and $t_{s,\text{DMC}}$ in Fig.~\ref{Fig:BER_vs_velocity}. It could be understood that by increasing the drift velocity in y axis from $\text{1}~\mu\text{m/s}$ for $W_x = \text{10}~\mu\text{m/s}$ and $t_{s,\text{DMC}} = \text{4.2} \text{ms}$ to $\text{100}~\mu\text{m/s}$, BER is improved from around $\text{10}^{\text{-6}}$ to around $\text{10}^{\text{-16}}$. It is also shown that by increasing $t_{s,\text{DMC}}$ from $\text{4.2} \text{ms}$ to $\text{4.4} \text{ms}$, the BER for fixed value of $W_y = \text{50}~\mu\text{m/s}$ is increased from $\text{2} \times \text{10}^{\text{-10}}$ to $\text{2} \times \text{10}^{\text{-7}}$. It is due to the fact that in 3-D model of environment, increasing symbol duration can not necessarily increase the reception probability \cite{bhatnagar20193,yilmaz}.
	
			\begin{figure}[]
		\centering
		\scalebox{.1}{}
		\includegraphics[width=330pt]{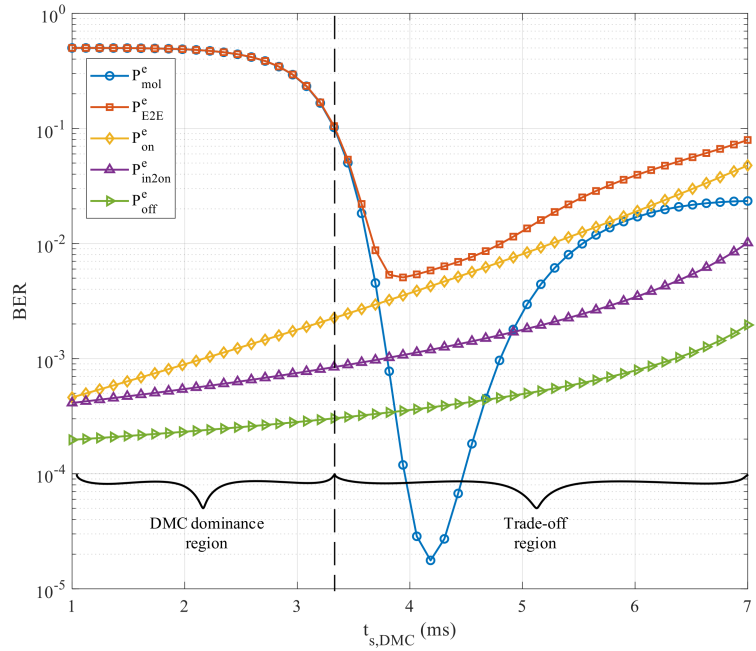}
		\caption{The E2E BER performance as a function of different values of symbol duration($W_x = \text{10}\mu\text{m/s}$, $W_y = \text{30} \mu\text{m/s}$, $W_z = \text{20}~\mu\text{m/s}$, $Z_x = \text{200}~\mu\text{m} = 2~G_x$, $Z_y =\text{100}~\mu\text{m}$, $Z_z = \text{20}~\mu\text{m} = 2~G_z$, and $Q_A = Q_B = 1000$). }
		\label{Fig:BER_vs_time}
	\end{figure}
	
	The E2E BER as a function of symbol duration in DMC is analyzed in Fig.~\ref{Fig:BER_vs_time}. Furthermore, the error probability for DMC, on-body, in2on, and off-body are illustrated. It could be shown that for $t_{s,\text{DMC}} < \text{3.327 ms}$, BER in DMC is dominant, and for $t_{s,\text{DMC}} > \text{3.327 ms}$ there is a trade-off between EC and DMC. However, the minimum E2E BER is reached when the trade-off between DMC and EC is achieved. The regions of DMC dominance and trade-off between EC and DMC are shown in Fig.~\ref{Fig:BER_vs_time}.
	
	\begin{figure}[]
		\centering
		\scalebox{.1}{}
		\includegraphics[width=330pt]{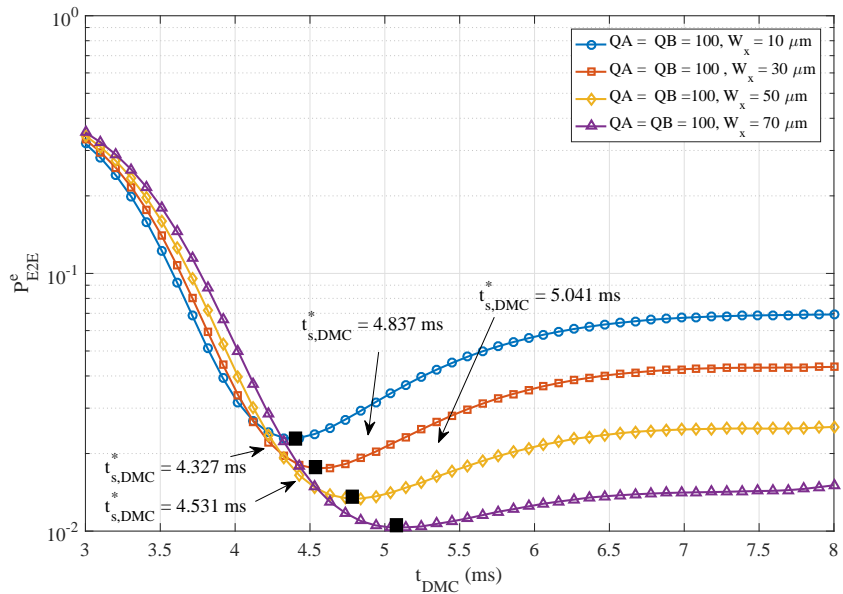}
		\caption{The performance of the proposed optimization problem to find the optimum $t^*_{s,\text{DMC}}$ ($W_x = \text{10}\mu\text{m/s}$, $W_y = \text{30} \mu\text{m/s}$, $W_z = \text{20}~\mu\text{m/s}$, $Z_x = \text{200}~\mu\text{m} = 2~G_x$, $Z_y =\text{100}~\mu\text{m}$, and $Z_z = \text{20}~\mu\text{m} = 2~G_z$).}
		\label{Fig:BER_vs_MAX}	
	\end{figure}

	The performance of the proposed optimization problem for different values of the drift velocity in x axis and fixed value of $Q_A = Q_B = \text{100}$ is shown in Fig.~\ref{Fig:BER_vs_MAX}. It is understood that by increasing the drift velocity, the minimum value of E2E BER is reached. However, in this case, the symbol duration is increased. It is due the fact that in such scheme,  increasing the velocity in x axis can divert the molecules from their direction to nodes R and D, to another direction. Therefore, the molecules need more times to arrive at the relay and destination nodes. By considering $W_x=~\text{10}~\mu\text{m/s}$, the optimum symbol duration to reach the minimum value of $P^e_{\text{E2E}} = \text{2}\times\text{10}^{\text{-2}}$ is 4.327~ms. In the scenario of considering $W_x = \text{70}~\mu\text{m/s}$, the optimum symbol duration is 5.041~ms, in which the minimum E2E BER is improved to $\text{1}\times\text{10}^{\text{-2}}$.

		\section{Conclusion} \label{Sec:Conclusion}
	In this paper, we investigated the E2E communication link consisting of the electromagnetic and molecular communication. First, we derived a closed-form expression for the E2E bit error probability of concatenation of molecular and wireless electromagnetic communications. Then, we formulated the optimization problem that aims at minimizing the E2E bit error probability of  the system to determine the optimal symbol durations for both molecular and wireless electromagnetic communications. In addition, we studied the impact of the parameters consisting of the detection threshold at the receiver, the location of the relay node, drift velocity, and symbol duration in MC on the performance. The results reveal that an adaptive system must be considered to achieve the minimum bit error rate and optimal performance for the E2E system.

\bibliographystyle{IEEEtran}
\bibliography{Tactile}

\end{document}